\documentclass[aps,pra,twocolumn]{revtex4-2}

\usepackage{bbm,color,amsmath,amsfonts,amsthm,graphicx}
\usepackage[colorlinks=true]{hyperref}
\usepackage{braket}
\usepackage{xcolor}
\usepackage{cancel}
\newcommand{\ita}[1]{{\em{#1}}}
\newcommand{\qUCCSD}{q-UCCSD}

\begin{document}

\title{Quantum simulation of electronic structure with a transcorrelated Hamiltonian: improved accuracy with a smaller footprint on the quantum computer}

\author{Mario Motta}
\affiliation{IBM Quantum, IBM Research Almaden, 650 Harry Road, San Jose, CA 95120, USA}
\author{Tanvi P. Gujarati}
\affiliation{IBM Quantum, IBM Research Almaden, 650 Harry Road, San Jose, CA 95120, USA}
\author{Julia E. Rice}
\affiliation{IBM Quantum, IBM Research Almaden, 650 Harry Road, San Jose, CA 95120, USA}
\author{Ashutosh Kumar}
\affiliation{Department of Chemistry, Virginia Tech, Blacksburg, VA 24061, USA}
\author{Conner Masteran}
\affiliation{Department of Chemistry, Virginia Tech, Blacksburg, VA 24061, USA}
\author{Joseph A. Latone}
\affiliation{IBM Quantum, IBM Research Almaden, 650 Harry Road, San Jose, CA 95120, USA}
\author{Eunseok Lee}
\affiliation{Mercedes-Benz Research and Development North America, Sunnyvale, CA 94085, USA}
\author{Edward F. Valeev}
\affiliation{Department of Chemistry, Virginia Tech, Blacksburg, VA 24061, USA}
\author{Tyler Y. Takeshita}
\affiliation{Mercedes-Benz Research and Development North America, Sunnyvale, CA 94085, USA}

\begin{abstract}
Quantum simulations of electronic structure with a transformed Hamiltonian that includes some electron correlation effects are demonstrated. The transcorrelated 
Hamiltonian used in this work is efficiently constructed classically, at polynomial cost, by 
an approximate similarity transformation with an explicitly correlated two-body unitary operator.
This Hamiltonian is Hermitian, includes no more than two-particle interactions, and is free of electron-electron singularities. 
We investigate the effect of such a transformed Hamiltonian on the accuracy and computational cost of quantum simulations by focusing on a widely used solver for the Schr\"odinger equation, namely the variational quantum eigensolver method, based on the unitary coupled cluster with singles and doubles (q-UCCSD) Ansatz. Nevertheless, the formalism presented here translates straightforwardly to other quantum algorithms for chemistry.
Our results demonstrate that a transcorrelated Hamiltonian, paired with extremely compact bases, produces explicitly correlated energies comparable to those from much larger bases.
For the chemical species studied here, explicitly correlated energies based on an underlying 6-31G basis had cc-pVTZ quality. 
The use of the very compact transcorrelated Hamiltonian reduces the number of CNOT gates required to achieve cc-pVTZ quality by up to two orders of magnitude, and the number of qubits by a factor of three.
\end{abstract}

\maketitle

\section{Introduction}

The simulation of quantum many-body systems is an important application for a quantum computer \cite{feynman1999simulating,lloyd1996universal,somma2003quantum,georgescu2014quantum,berry2015simulating,low2019hamiltonian,childs2018toward}.
In the context of quantum chemistry and materials science, a key example of such an application is 
the electronic structure (ES) problem, namely solving for the ground or low-lying eigenstates 
of the electronic Schr\"{o}dinger equation for atoms, molecules, and materials.
In recent years, a variety of quantum algorithms has delivered promising results  in the calculation 
of potential energy curves, ground- and excited-state energies and ground-state correlation functions 
for molecules comprising first and second row elements \cite{o2016scalable,kandala2017hardware,cao2019quantum,grimsley2019adaptive,rice2020quantum,parrish2019quantum,stair2019multireference,huggins2020non}.

Despite the rapid development of quantum hardware and algorithms, modern quantum computation platforms 
are immature. This fact, combined with the limitations of classical simulators and popular one-to-one
mappings of spin-orbitals to qubits, has resulted in most quantum ES simulations reported to date
employing minimal basis sets (i.e. describing core and valence orbitals only) or being restricted to 
active spaces of a few orbitals and electrons. While simulations based on minimal basis sets and/or 
small active spaces continue to provide benchmarks, useful quantum simulations will require significant quantum resources. Today routine 
classical ES calculations may contain hundreds to thousands of basis functions that would need to be mapped
to logical qubits. Thus, it is clear we need approaches that can give the desired accuracy with fewer quantum resources.

Two such approaches are currently being explored. 
One approach is to perform small  calculations on the quantum
computer followed by classical post-processing to partially correct for basis set errors associated with using too few
qubits \cite{takeshita2020increasing}. 
The second is to reduce the quantum resources
required for more accurate calculations (measured in the number of qubits and quantum gates). In this 
paper, we focus on the latter approach.

The conventional description of the many-body wave function as a superposition of single Slater 
determinants offers a natural and efficient way to address static
electronic correlation. However, it does not treat dynamic correlation efficiently, which is
necessary to achieve chemical accuracy, as compared to experiment. The inefficient treatment of dynamic correlation leads to slow
convergence to the complete basis set (CBS) limit and thus requires the use of large basis sets to attain such accuracy.

Indeed, due to the Coulomb singularity of the electronic interaction, the short-range 
\ita{dynamical correlation} introduces cusps \cite{kato1957eigenfunctions,Pack:1966fw,Kutzelnigg:1992kj} 
at the points of coalescence between two electrons. These cusps cannot be approximated efficiently 
by orbital product expansions and require explicit parametric dependence of the wave function on 
the inter-electronic distances. Although the use of such \ita{explicitly correlated} wave functions 
has been commonplace for high precision computations of small systems since the pioneering work of 
Hylleraas in 1929 \cite{Hylleraas:1929jb}, \ita{efficient} application of explicitly correlated methods 
to molecules has become possible only due to the development of the ideas proposed by Kutzelnigg
\cite{Kutzelnigg1985R12}. The explicitly correlated F12 (originally known as ``R12") methods 
dramatically improve the convergence of the electronic energy and other molecular properties with 
respect to the basis set size. 
Numerous improvements over the years
\cite{Klopper2002F12Aux,Manby2003F12DF,TenNo2004F12Quad,Tenno2004F12STG,valeev2004improving,Hattig2005CCSDR12,Werner2007MP2F12,Valeev2008F12,watson2016correct} 
have now made the F12 calculations quite black-box and robust \cite{Kong:2012dx,Hattig:2012dz,TenNo:2012bb}.

In this work, we consider the use of explicit correlation for defining a similarity-transformed
Hamiltonian that includes the dynamical electron correlation effects following the recipe of Yanai 
and Shiozaki for canonical transcorrelated F12 (CT-F12) Hamiltonian \cite{yanai2012canonical}.
The CT-F12 theory can be seen as an extension of the transcorrelated Hamiltonian approach originally introduced by Boys and Handy \cite{BoysHandy1969transcorrelated} and later improved by Ten-no \cite{Tenno2000transcorrelated} and Luo \cite{Luo2010transcorrelated}, where singularity-free 
Hamiltonians are constructed from the similarity transformation of the original Hamiltonian through a geminal correlation operator $\hat{A}$,
\begin{equation}
\hat{H} \to \hat{H}^\prime = e^{-\hat{A}} \hat{H} e^{\hat{A}} \quad .
\label{eq:ST}
\end{equation}
What makes the CT-F12 method robust and simpler to use, compared to the earlier transcorrelated 
Hamiltonian formalisms, is the choice of the unitary operator in the similarity transformation, $e^{\hat{A}}$, (where $\hat{A}=-\hat{A}^{\dagger}$), thereby ensuring that the effective 
Hamiltonian remains Hermitian, and in the truncation of the approximate Baker–Campbell–Hausdorff
(BCH) expansion of Eq.~\eqref{eq:CT} to include only 1 and 2-body effective Hamiltonian elements, 
following the ideas from the canonical transformation (CT) method \cite{Chan2006CanonicalTransform,Chan2010CanonicalTransform},
\begin{equation}
\begin{split}
\hat{H}^\prime &= e^{ - \hat{A} } \hat{H} e^{ \hat{A} } \\
&\approx \hat{H} + {[ \hat{H}, \hat{A}]}_{1,2} + \frac{1}{2} {{[[ \hat{H}, \hat{A}]}_{1,2}, \hat{A}]}_{1,2} + \dots \; .
\label{eq:CT}
\end{split}
\end{equation}
where, $[..]_{1,2}$ refers to the retention of only 1 and 2-body elements of the given commutator. 
The operator $\hat{A}$ is defined using the Slater-type geminal, 
$\hat{F}_{12}(r_{12})= -\gamma^{-1} \, e^{-\gamma r_{12}}$, where the inverse length scale $\gamma$ 
is commensurate with the correlation length scale of the valence electrons and in practice is tuned 
for a given orbital basis set \cite{Klopper2005Gamma}.
Only the pure two-body (de)excitation component (relative to a zeroth-order reference) is included 
in $\hat{A}$, and the geminal is scaled by \{1/2,1/4\} when acting on \{singlet,triplet\} electron 
pairs in accordance with the spin dependence of the electron-electron cusp \cite{Pack:1966fw} (this 
is the so-called SP Ansatz of Ten-no \cite{TenNo2004SPAnsatz,Zhang:2012it}).
Thus, the exact form of the operator is known \textit{a priori}, albeit the operator introduces a 
dependence on the particular reference and the geminal length scale.

In the present work the CT-F12 Hamiltonian is used in conjunction with the variational quantum 
eigensolver (VQE) method \cite{peruzzo2014variational,mcclean2016theory,wang2019accelerated,bauer2020quantum}. To the best of our knowledge, this is the first study to combine explicitly correlated techniques 
with quantum algorithms, to achieve higher accuracy without increasing quantum resources (e.g. number of qubits needed to represent the Hamiltonian).
A slightly later contribution \cite{mcardle2020improving} considered a different transcorrelated method, which is characterized by a non-hermitian Hamiltonian, in combination with variational imaginary-time evolution techniques. The results of the two works are thus complementary, and highlight the importance of exploring different transcorrelated approaches for quantum simulation.

We study several chemical species comprising hydrogen (H$_2$, H$_3^+$) and closed-shell, first-row 
hydrides (LiH, BH, HF) using Pople \cite{ditchfield1971self,hariharan1973influence} and correlation-consistent  \cite{dunning1989gaussian} basis sets, while adopting the well-established 
unitary coupled cluster with singles and doubles (q-UCCSD) 
Ansatz \cite{kutzelnigg1982quantum,kutzelnigg1983quantum,kutzelnigg1985quantum,cooper2010benchmark}.

In order to focus on the CT-F12 method, we use the VQE method and q-UCCSD Ansatz since these latter techniques are now part of the standard toolkit of quantum simulation. However, it should be noted the Hermitian nature and the compact form of the CT-F12 Hamiltonian studied makes its integration with other Ans\"{a}tze and quantum algorithms very straightforward.

In published literature, CT-F12 methods have been used to extrapolate from reasonably sized basis 
sets to much larger basis sets \cite{yanai2012canonical}. In this work, motivated by the desire to 
fit the budget of contemporary quantum hardware, we investigated extrapolation from small basis 
sets (e.g. 6-31G) to somewhat larger basis sets. Note that this is not a direct translation from 
the classical CT-F12 algorithms.

The remainder of the present work is structured as follows: the CT-F12 and VQE methods are briefly 
reviewed in section \ref{sec:methods}, results are presented in section \ref{sec:results}, 
conclusions are drawn in section \ref{sec:conclusions}.

\section{Methods}
\label{sec:methods}

\begin{figure}[h!]
\centering
\includegraphics[width=0.95\columnwidth]{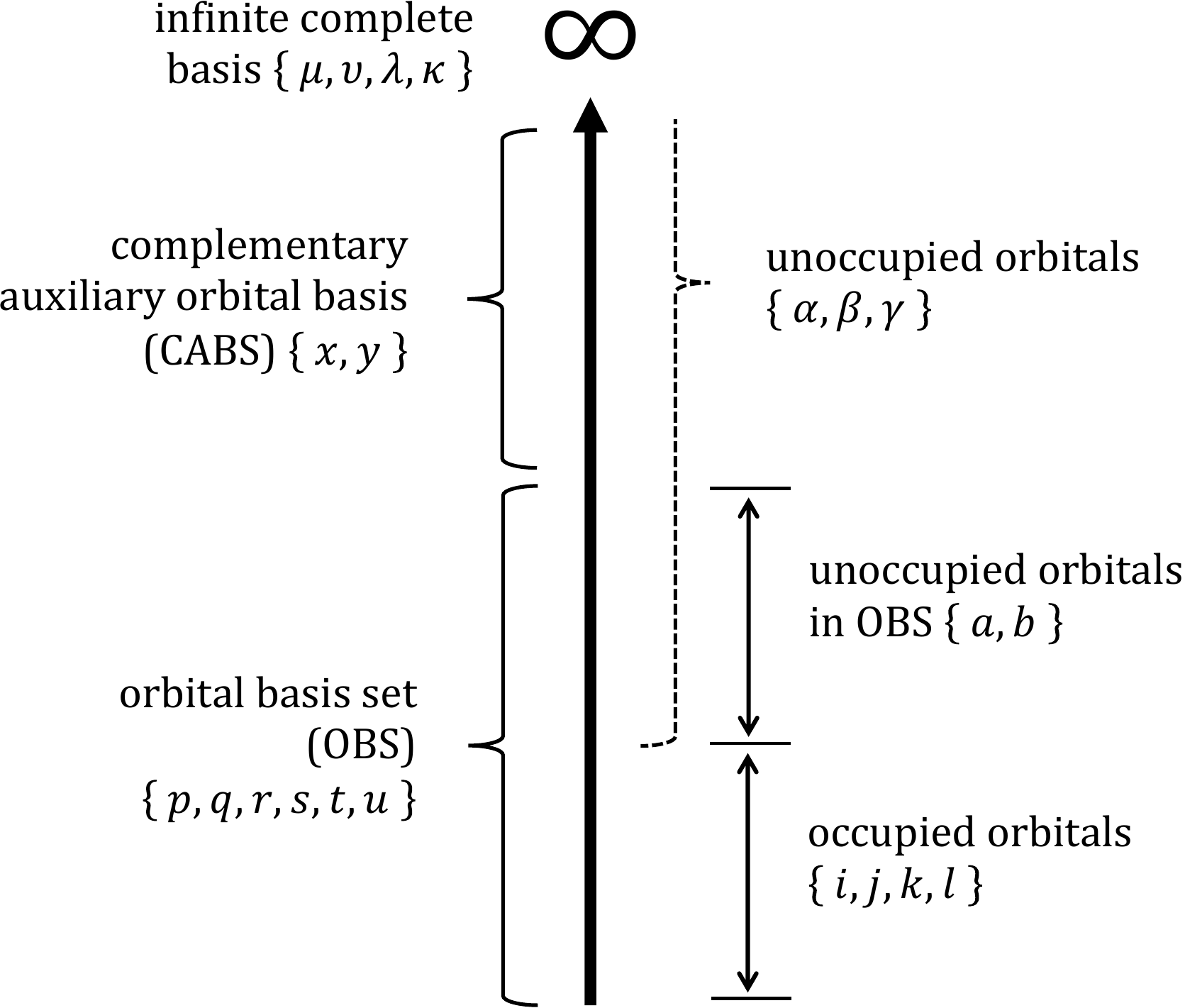}
\caption{
Schematic notation of orbital indices in the CT-F12 method.
Reprinted from \cite{yanai2012canonical}, with the permission of AIP Publishing.
}
\label{fig:0}
\end{figure}

\subsection{Canonical transcorrelated F12 Hamiltonian}
In the CT-F12 method, two main approximations are employed in addition to the approximate BCH expansion 
of Eq.~\eqref{eq:CT}: (a) the expansion is truncated to only include up to double commutators 
and (b) in the double commutator term, the full Hamiltonian $\hat{H}$ is replaced by its 
effective 1-body constituent, the Fock operator $\hat{F}$,
\begin{equation}
\hat{H}^\prime 
\approx 
\hat{H} 
+ {[ \hat{H}, \hat{A}]}_{1,2} 
+ \frac{1}{2} {{[[ \hat{F}, \hat{A}]}_{1,2}, \hat{A}]}_{1,2} 
\quad.
\label{eq:CT-F12}
\end{equation}
These approximations are consistent with the ones employed in some approximate CT-F12 theories \cite{Tew2010CCF12} and ensure that the effective Hamiltionian is correct through second-order 
in the perturbation (in the M\o{}ller-Plesset sense). 
Of course, since the unitary transformation $e^{-\hat{A}}$ is applied approximately, CT-F12 energies are not guaranteed to be variational, especially in multireference situations, where high-order contributions are important.

Figure~\ref{fig:0} refers to the notation of orbital indices used from  \cite{yanai2012canonical}.
The molecular Hamiltonian in spin-free form is written as
\begin{equation}
\hat{H} = h^\mu_\nu \hat{E}^\nu_\mu + \frac{1}{2} g^{\mu\lambda}_{\nu\kappa}
\hat{E}^{\nu\kappa}_{\mu\lambda}
\quad,
\end{equation}
where indices $\kappa$, $\lambda$, $\mu$, $\nu$ label formal basis
in the (complete) 1-particle Hilbert space, $h^\mu_\nu$ and
$g^{\mu\lambda}_{\nu\kappa}$ are matrix elements of the one- and two-body parts of the Hamiltonian,
\begin{align}
h^\mu_\nu = & \bra{\nu}\hat{H}_1\ket{\mu} \;, \\
g^{\mu\lambda}_{\nu\kappa} = & \bra{\nu \kappa}\hat{H}_2\ket{\mu \lambda} \;.
\end{align}
Operators
\begin{equation}
\hat{E}^\nu_\mu = \sum_{\sigma = \uparrow\downarrow}
\hat{c}^\dagger_{\nu\sigma} \hat{c}_{\mu\sigma}
\;,\;
\hat{E}^{\nu\kappa}_{\mu\lambda}
=
\sum_{\sigma\tau = \uparrow\downarrow}
\hat{c}^\dagger_{\nu\sigma} 
\hat{c}^\dagger_{\kappa\tau} \hat{c}_{\lambda\tau}
\hat{c}_{\mu\sigma} \;,
\end{equation}
are the spin-summed transition operators composed of the traditional creators/annihilators $\hat{c}^\dagger_p/\hat{c}_q$. In all the equations, Einstein summation convention
is implied. The Fock operator is written as
\begin{equation}
\hat{F} = f^\mu_\nu \hat{E}^\nu_\mu
\;,\;
f^\mu_\nu = h^\mu_\nu + \rho_\lambda^\kappa 
\left( g^{\mu\lambda}_{\nu \kappa} - \frac{1}{2} g^{\lambda\mu}_{\nu\kappa} \right)
\;,
\end{equation}
where $\rho$ is the one-body density matrix at the Hartree-Fock level.
The orbital basis (OBS) $p$,$q$,$r$,$s$,$t$,$u$ is divided into occupied $i$,$j$,$k$,$l$ and 
unoccupied $a$,$b$ parts. The orbitals of the complete basis set (CBS) are represented by
$\mu$,$\nu$,$\lambda$,$\kappa$ with the unoccupied ones denoted by $\alpha$,$\beta$,$\gamma$. 
Finally the complementary auxiliary orbital basis set (CABS) \cite{valeev2004improving},
is denoted by $x$,$y$.

As mentioned before, $\hat{A}$ is an anti-hermitian operator,
\begin{equation}
\hat{A} = \frac{1}{2} G^{\alpha\beta}_{ij}
\left( \hat{E}^{\alpha\beta}_{ij} - \hat{E}^{ij}_{\alpha\beta} \right)
\;,
\label{eq:A}
\end{equation}
where
\begin{equation}
G^{\alpha\beta}_{ij} = 
\frac{3}{8}
\,
\langle \alpha \beta | \hat{Q}_{12} \hat{F}_{12} | ij \rangle
+ 
\frac{1}{8}
\,
\langle \alpha \beta | \hat{Q}_{12} \hat{F}_{12} | ji \rangle,
\;
\label{eq:G}
\end{equation}
is defined in terms of a geminal (2-body correlator)
\begin{equation}
\hat{F}_{12}(r_{12}) = - \frac{ e^{-\gamma r_{12}} }{\gamma},
\end{equation}
and a projector ensuring orthogonality to the unoccupied orbital products $\ket{ab}$,
\begin{equation}
\hat{Q}_{12} = 
1 -
\hat{V}_1
\hat{V}_2
\;,
\end{equation}
where $\hat{V}_i$ projects the $i$-th particle state onto the unoccupied orbitals represented 
in the orbital basis set. Since our work deals with the unitary coupled cluster method with a 
Hartree-Fock reference, the strong orthogonality (i.e. pure 2-body character) of the geminal 
is automatically ensured by the form of the operator $\hat{A}$ in Eq.~\eqref{eq:A}. 

The coefficients 3/8 and 1/8 in Eq.~\eqref{eq:G} arise from the spin-dependent cusp condition 
coefficients. \cite{TenNo2004SPAnsatz,Zhang:2012it}
Since optimized values of the correlation factor $\gamma$ are available in the literature only 
for standard medium and large sized basis sets \cite{Klopper2005Gamma}, we chose those values
of $\gamma$ which for a given molecule and basis set, gave the lowest CT-F12/CCSD energies at 
the equilibrium geometry. Table ~\ref{tab:1} lists the values of $\gamma$ used for the 6-31G 
and cc-pVDZ basis sets for different molecules.

\begin{table}[h!]
\centering
\begin{tabular}{ccccc} 
\hline\hline
Molecule & 6-31G & cc-pVDZ \\
\hline
H$_2$   & 0.7 & 0.7 \\
H$_3^+$ & 0.7 & 0.7 \\
LiH     & 0.6 & 0.6 \\
BH      & 0.7 & 0.7 \\
HF      & 1.3 & 1.3 \\
\hline\hline
\end{tabular}
\caption{Optimized values of the correlation factor $\gamma$ for each molecule and basis set}
\label{tab:1}
\end{table}

Finally, the transformed Hamiltonian takes the form
\begin{equation}
\hat{H}^\prime = \overline{h}^p_q \hat{E}^q_p + \frac{1}{2} \overline{g}^{pr}_{qs}
\hat{E}^{qs}_{pr}
\;,
\end{equation}
where the explicit formulas for one and two body elements are shown in \cite{yanai2012canonical}.
The overall complexity of computing the transformed Hamiltonian for the Hartree-Fock reference is
$\mathcal{O}(N^6)$; the cost grows quadratically with the CABS basis rank when approach C of 
reference \cite{Kedzuch:2005hr} is used to compute the geminal matrix element of the Fock operator, 
but this cost can be robustly lowered further to linear \cite{Pavosevic:2016bc}.
Note that the Hamiltonian $\hat{H}^\prime$ is Hermitian, only contains one- and two- body terms,
and since its two-body part is not multiplicative, it has lower symmetry than the original 
Hamiltonian (e.g., $\overline{g}^{pr}_{qs} \neq \overline{g}^{ps}_{qr}$ (for $\hat{H}^\prime$)  whereas $g^{pr}_{qs} = g^{ps}_{qr}$) (for $\hat{H}$).
Due to technical limitations, Yanai and Shiozaki symmetrized the 2-body part of the transcorrelated Hamiltonian [($\overline{g}^{pr}_{qs} + \overline{g}^{ps}_{qr})/2 \rightarrow \overline{g}^{pr}_{qs}$] 
to obtain the same symmetry as the original Hamiltonian \cite{yanai2012canonical}, however no such
symmetrization was performed here.

cc-pVDZ-F12-OptRI basis set \cite{Peterson2010CABS} was used as our 
CABS basis set utilizing the CABS+ approach \cite{valeev2004improving} in all the reported calculations.
Finally, evaluation of the CT-F12 Hamiltonian was implemented through the ``plugout" feature of the 
C++ based MPQC4 software package \cite{MPQCPublicRepo} i.e. the MPQC4 toolkit was imported as a library 
in an external C++ program.

\subsection{The variational quantum eigensolver}

Variational quantum state preparation algorithms are a class
of quantum algorithms, that have been conjectured to be particularly amenable to near-term quantum devices.
In close analogy with classical variational approaches, one chooses a class of Ansatz states 
approximating the ground state of the Hamiltonian of interest. 
In general, such an Ansatz is defined by an initial state $| \Psi_0 \rangle$ and a unitary circuit $\hat{U}(\theta)$ defined by a set of classical variational parameters
$\theta \in \Theta$, leading to a family $| \Psi(\theta) \rangle = \hat{U}(\theta) | \Psi_0 \rangle$ 
of wavefunctions.
For each state $| \Psi(\theta) \rangle$, the energy 
$E(\theta) = \langle \Psi(\theta) | \hat{H} | \Psi(\theta) \rangle$ 
provides an upper bound to the ground-state energy, and the parameters $\theta$ can be optimized to 
lower the energy of the state $| \Psi(\theta) \rangle$ relying on a classical optimization algorithm.
This procedure defines the variational quantum eigensolver or VQE method \cite{peruzzo2014variational}.

The choice of the variational family $\{ | \Psi(\theta) \rangle \}_{\theta}$ is motivated by a 
combination of factors. On the one hand, it is important to produce an accurate approximation 
to the true ground state of the system, to offer chemically meaningful results. Secondly, the 
optimization problem of minimizing $E(\theta)$ as a function of the parameters $\theta$ has to 
be well-behaved, to give the ability of finding energy minima. 
Finally, for calculations on quantum hardware, it is important 
to have circuits that fit their budget of available gates, qubit connectivity and coherence times of contemporary quantum hardware.

The diversity of problems investigated in quantum simulation and the ever-changing capabilities 
of quantum hardware have motivated a large variety of proposals in recent years, see for example
\cite{kandala2017hardware,grimsley2019adaptive,kim2017robust,liu2019variational,schon2005sequential},
making the design and benchmark of variational quantum Ans\"{a}tze an active area of research.

\subsection{Unitary coupled cluster with singles and doubles}

An important example of a variational family suggested for applications in quantum chemistry is 
the unitary coupled cluster (UCC) Ansatz
\cite{kutzelnigg1982quantum,kutzelnigg1983quantum,kutzelnigg1985quantum,cooper2010benchmark,romero2018strategies},
\begin{equation}
\label{eq:ucc}
\begin{split}
&| \Psi_{\mbox{UCC}}(\theta) \rangle = e^{ \hat{T} - \hat{T}^\dagger } | \Psi_0 \rangle
\quad, \\
&\hat{T} = \sum_{k=1}^d \sum_{ \substack{i_1 \dots i_k \\ a_1 \dots a_k} } 
\theta^{a_1 \dots a_k}_{i_1 \dots i_k} \,
\hat{c}^\dagger_{a_1} \dots \hat{c}^\dagger_{a_k}
\hat{c}_{i_1} \dots \hat{c}_{i_k} \quad,
\end{split}
\end{equation}
where $| \Psi_0 \rangle$ denotes the Hartree-Fock state, $d$ denotes the maximum order of 
excitations in the UCC wavefunction, and the cluster amplitude tensors 
$\theta^{a_1 \dots a_k}_{i_1 \dots i_k}$ are antisymmetric in the indices $a_1 \dots a_k$ 
and $i_1 \dots i_k$. In particular, $d=2$ in Eq.~\eqref{eq:ucc} gives unitary coupled cluster 
with single and double excitations (UCCSD).

This choice of Ansatz is very natural in situations where mean-field theory is successful, 
which suggests that excitations relative to the mean-field state $| \Psi_0 \rangle$ in the 
actual ground state wavefunction should be small, or equivalently that dynamical correlation 
dominates the problem.

Standard coupled cluster Ansatz $e^{ \hat{T} } | \Psi_0 \rangle$ is widely used in classical 
quantum chemistry but is challenging to implement on a quantum device due to the non-unitarity 
of $e^{ \hat{T} }$, whereas the converse is true for UCC.
Understanding the relationship between standard and unitary coupled cluster Ansatz\"{e} is an 
active area of research \cite{cooper2010benchmark,sokolov2019quantum}, of value to both chemistry 
and quantum information science. 
To be able to implement the UCCSD ansatz on the quantum computer, a Trotter decomposition step 
as explained in Section \ref{sec:D} is used. As per the nomenclature adopted in previous literature \cite{barkoutsos2018quantum,rice2020quantum}, we refer to this Ansatz as \qUCCSD.

\section{Results}
\label{sec:results}

\begin{figure*}
\centering
\includegraphics[width=0.6\textwidth]{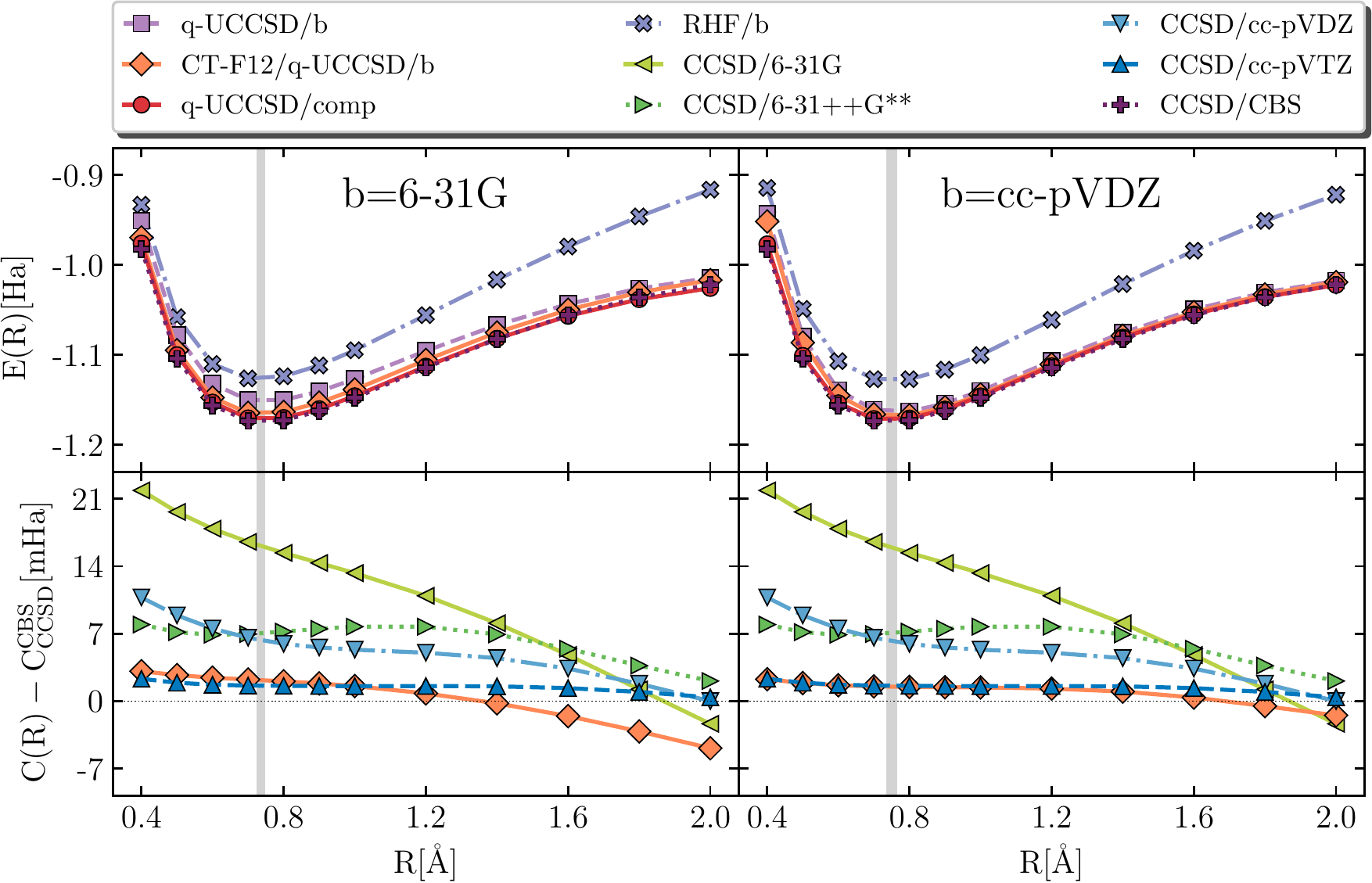} 
\caption{Top: 
Potential energy curves for H$_2$ from RHF, q-UCCSD, CT-F12/q-UCCSD, q-UCCSD/comp and CCSD/CBS, using the 6-31G (left) 
and cc-pVDZ (right) bases.
Bottom: Comparison between classical CCSD/CBS correlation energies and classical CCSD/(6-31G, 6-31++G$^{**}$, cc-pVDZ, cc-pVTZ), CT-F12/q-UCCSD/6-31G correlation energies (left).
Comparison between classical CCSD/CBS correlation energies and classical CCSD/(6-31G, 6-31++G$^{**}$, cc-pVDZ, cc-pVTZ), CT-F12/q-UCCSD/cc-pVDZ correlation energies (right).
Lines are a guide for the eye, and gray bands represent the 
range of computed equilibrium bond lengths.
}
\label{fig:1}
\end{figure*}

The calculations performed in this work involved initial pre-processing by quantum chemistry codes (in this case MPQC4 and PySCF) \cite{sun2018pyscf,sun2020recent,MPQCPublicRepo}) on classical
computers, to generate optimized mean-field orbitals and matrix elements of the regular and explicitly
correlated Hamiltonian prior to performing computations with quantum simulators. 
The restricted Hartree-Fock (RHF) singlet state was chosen as the initial state for all of the 
calculations described here. All correlated calculations used the frozen core approximation. 
It is worth observing that the frozen core approximation not only economizes simulations by 
removing orbitals and electrons, but is also justified by the nature of the basis sets used 
in the present work, since they are constructed for valence-only correlated calculations.

Having selected a set of single-electron orbitals for each of the studied species, VQE computations 
were performed with quantum simulators. We used IBM's open-source library for quantum computing, 
Qiskit \cite{aleksandrowicz2019qiskit}.
Qiskit Aqua contains implementations of techniques to map the fermionic Fock space onto the Hilbert 
space of a register of qubits, and an implementation of the VQE algorithm.
Here we use the tapering-off technique \cite{bravyi2017tapering,setia2019reducing} to account for 
molecular point group symmetries and reduce the number of qubits required for a simulation. 
In analogy with conventional symmetry-adapted quantum chemistry calculations, this reduction does 
not introduce additional approximations in the calculations. In the VQE simulations, we used the 
quantum circuit defined in \cite{barkoutsos2018quantum} to implement the \qUCCSD ~ Ansatz.

We then minimized the expectation value of the Hamiltonian with respect to the parameters in the 
circuit. The minimization was carried out using the classical optimization method, L-BFGS-B \cite{zhu1997algorithm,morales2011remark}. 
We ran our experiments on the statevector simulator of Qiskit.

For the CT-F12 Hamiltonian, \qUCCSD\, \ita{correlation energies were computed as differences between 
total} CT-F12/q-UCCSD{\,} \ita{energies and} RHF \ita{energies with regular Hamiltonian}, as outlined in \cite{yanai2012canonical}.
For comparison with the F12 results, restricted, regular coupled cluster with singles and doubles 
(CCSD) calculations were performed using PySCF.
CBS energies are computed extrapolating cc-pVxZ (x=2,3,4,5) RHF energies with the formula $E_{\mathrm{RHF},x} = \alpha + \beta e^{\gamma x}$, and cc-pVxZ (x=3,4,5) correlation energies with the formula $C_x = \alpha^\prime + \frac{\beta^\prime}{x^3}$ following \cite{dunning1989gaussian}.

In addition to that, we list the energies of a composite method, where the Hartree-Fock energy is calculated with a large basis set (namely, cc-pVTZ) using the regular Hamiltonian, 
and added to the CT-F12/q-UCCSD~correlation energies (namely determined using the CT-F12 Hamiltonian
and a smaller basis set, 6-31G unless otherwise specified). 
The composite approach removes the effect of basis set incompleteness both at one-body (Hartree-Fock energy) and two-body level (dynamic correlation energy). As such, the composite approach consistently yields the best properties reported in this work.

Such a composite method is well suited for a hybrid classical/quantum methodology. The Hartree-Fock 
procedure, which in its canonical formulation scales at most as $N^4$, is appropriate for the 
classical hardware, whereas the calculation of the correlation energy, which can cost as much 
as $2^N$, is best mapped to the quantum computer.

For the sake of compactness, we adopt the following notation: standard calculations are denoted
by method/basis (e.g. q-UCCSD/6-31G), explicitly correlated calculations by CT-F12/method/basis (e.g. 
CT-F12/q-UCCSD{/6-31G}), and composite methods by RHF/basis + correlated method (e.g. HF/cc-pVTZ + 
CT-F12/q-UCCSD{/6-31G}) or simply correlated method / comp. Note that CCSD and 
q-UCCSD are equivalent to full configuration interaction (within the same basis) for systems with two electrons.

We first present results for hydrogen \ref{sec:H2} and the trihydrogen cation \ref{sec:H3plus}, followed by results for some first row hydrides (LiH, BH and HF) in section \ref{sec:D}.

\subsection{Hydrogen molecule}
\label{sec:H2}

In Figure \ref{fig:1} we compute the potential energy surface of the hydrogen molecule using 
RHF, CCSD, q-UCCSD, and CT-F12-q-UCCSD with the 6-31G and cc-pVDZ basis sets.

As seen, the difference between the q-UCCSD and CT-F12-q-UCCSD energies is more pronounced 
when the underlying basis is 6-31G.
In the lower portion of Figure ~\ref{fig:1}, 
we compared q-UCCSD/6-31G and CT-F12/q-UCCSD/6-31G correlation energies against CCSD/6-31G,
CCSD/6-31G$^{**}$, CCSD/6-31++G and CCSD/6-31++G$^{**}$ correlation energies.
Note that the positive (or close to zero) correlation energy differences seen for the larger 
basis sets reflect that CT-F12/q-UCCSD/6-31G correlation energies have quality better than 
(or comparable to) the regular correlation energies for these larger basis sets.

\begin{table} 
\centering
\begin{tabular}{cccc}
\hline\hline
method/basis         & type    & $R_{eq} [\text{\AA}]$ & $\omega [\mathrm{cm}^{-1}]$ \\
\hline
RHF/6-31G            & regular & 0.7312(6) & 4660(42) \\
q-UCCSD/6-31G        & regular & 0.7468(5) & 4386(25) \\
q-UCCSD/6-31G        & CT-F12  & 0.7397(6) & 4462(29) \\
\hline
RHF/cc-pVDZ          & regular & 0.7488(7) & 4617(34) \\
q-UCCSD/cc-pVDZ      & regular & 0.7613(6) & 4414(22) \\
q-UCCSD/cc-pVDZ      & CT-F12  & 0.7572(6) & 4432(24) \\
\hline
CCSD/CBS             & regular & 0.740(1)  & 4439(59) \\
q-UCCSD/comp$^{(a)}$ & comp    & 0.7480(8) & 4314(44) \\
q-UCCSD/comp$^{(b)}$ & comp    & 0.7471(8) & 4332(35) \\
\hline\hline
\end{tabular}
\caption{RHF, CCSD and q-UCCSD equilibrium bond lengths and vibrational frequencies for H$_2$ at 6-31G and cc-pVDZ level with regular and CT-F12 Hamiltonians, and extrapolated to the CBS limit. 
Numbers in round brackets denote uncertainties from the fitting procedure. 
Experimental values are 
$R_{eq} = 0.741 \, \text{\AA}$ and $\omega = 4401 \, \mathrm{cm}^{-1}$ respectively \cite{johnson1999nist}.
"comp" refers to the composite RHF/cc-pVTZ + CT-F12/q-UCCSD/6-31G 
and RHF/cc-pVTZ + CT-F12/q-UCCSD/cc-pVDZ methods (a,b respectively).
}
\label{tab:2}
\end{table}

CT-F12/q-UCCSD/6-31G correlation energies have quality comparable to regular CCSD/6-31++G$^{**}$ 
correlation energies suggesting that, for split-valence basis sets \cite{ditchfield1971self,davidson1986basis},
explicit correlation accounts for the combined effect of polarization and diffuse functions.

\begin{figure*}
\centering
\includegraphics[width=0.75\textwidth]{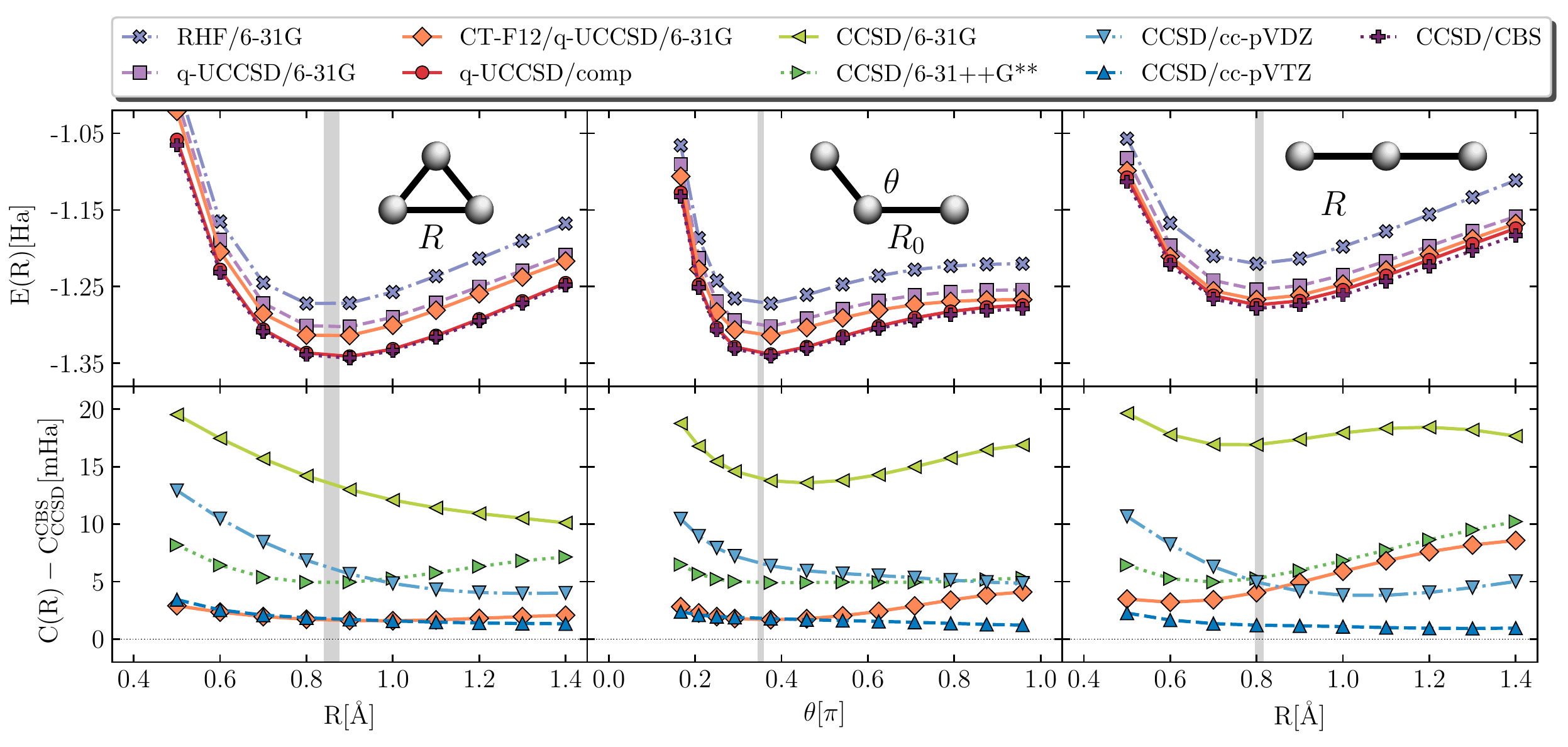} 
\caption{
Potential energy curves for H$^+_3$ 
from RHF/6-31G, q-UCCSD/6-31G, CT-F12/q-UCCSD/6-31G, q-UCCSD/comp and CCSD/CBS.
Bottom: Comparison between classical CCSD/CBS correlation energies and classical CCSD/(6-31G, 6-31++G$^{**}$, cc-pVDZ, cc-pVTZ), CT-F12/q-UCCSD/6-31G correlation energies.
Results are shown as a function of $R$ for the stretching of a triangular (left) and 
a linear (right) molecule, and for the variation in $\theta$ from the triangular to 
the linear conformer (middle).
Lines are a guide for the eye, gray bands represent the range of RHF and q-UCCSD 
equilibrium bond lengths, and sketches in the panels illustrate the meaning of the 
coordinates $R$ and $\theta$ with $R_0 = 0.81 \text{\AA}$.
}
\label{fig:2}
\end{figure*}
\begin{figure*}
\centering
\includegraphics[width=0.75\textwidth]{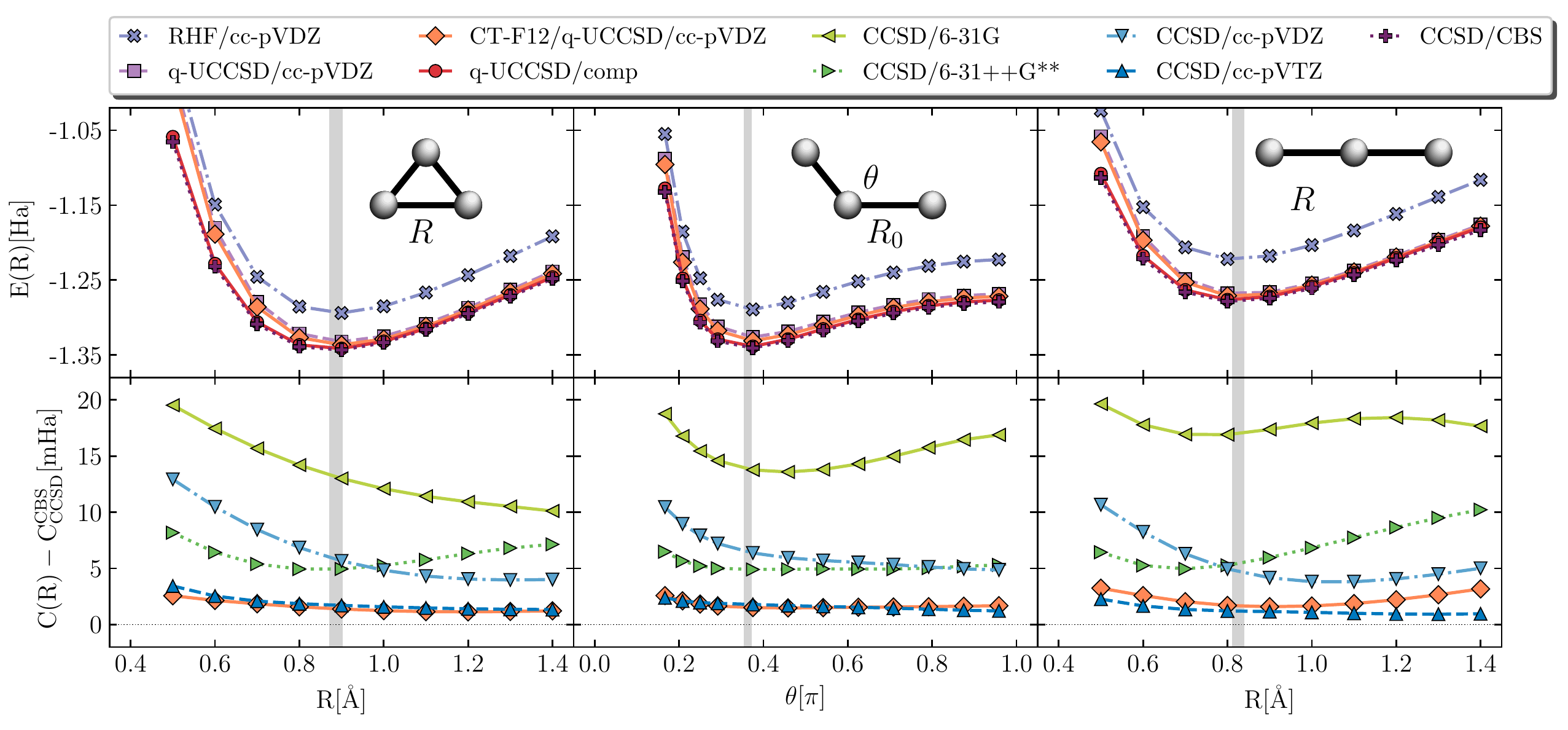}
\caption{
Potential energy curves for H$^+_3$ 
from RHF/cc-pVDZ, q-UCCSD/cc-pVDZ, CT-F12/q-UCCSD/cc-pVDZ, q-UCCSD/comp and CCSD/CBS.
Bottom: Comparison between classical CCSD/CBS correlation energies and classical CCSD/(6-31G, 6-31++G$^{**}$, cc-pVDZ, cc-pVTZ), CT-F12/q-UCCSD/cc-pVDZ correlation energies.
}
\label{fig:3}
\end{figure*}

Comparison between CT-F12/q-UCCSD/cc-pVDZ and regular CCSD/cc-pVxZ (x=D,T), CCSD/CBS correlation energies \cite{dunning1989gaussian} suggests that explicit correlation yields correlation energies of quality comparable with the next basis set in the series, cc-pVTZ. In the large $R$ regime, correlation energies are slightly overestimated.

Equilibrium bond lengths and vibrational frequencies, obtained by fitting the computed potential 
energy surfaces around the minimum to a Morse potential, are listed in Table \ref{tab:2}.
We observe that the composite RHF/cc-pVTZ + CT-F12/q-UCCSD/6-31G and 
RHF/cc-pVTZ + CT-F12/q-UCCSD/cc-pVDZ energies lead to equilibrium geometries and vibrational 
frequencies in good agreement with CCSD/CBS.

\subsection{Tri-hydrogen cation}
\label{sec:H3plus}

In Figures \ref{fig:2} and \ref{fig:3} we compute potential energy surfaces for the 
tri-hydrogen cation, using the 6-31G and cc-pVDZ bases, respectively.
We considered three conformers: 
(i) an equilateral triangle with variable bond length $R$, 
(ii) a linear geometry with variable bond length $R$, and 
(iii) an isosceles triangle with fixed bond length $R_0 = 0.81 \text{\AA}$ and variable angle $\theta$.

As seen in the lower portion of Figure \ref{fig:2}, CT-F12/q-UCCSD/6-31G correlation 
energies have quality superior to the CCSD/6-31G and 
CCSD/6-31++G$^{**}$ correlation energies.
In Figure \ref{fig:3}, CT-F12/q-UCCSD/cc-pVDZ correlation energies have quality 
comparable to CCSD/cc-pVTZ correlation energies, as seen above for $\text{H}_\text{2}$.
In both cases, CT-F12 correlation energies lie a few mHa above CBS correlation energies.

In Table \ref{tab:3}, we list the equilibrium bond lengths for the linear and equilateral 
triangle conformers, and the energy difference between them. We observe that both q-UCCSD 
and CT-F12/q-UCCSD predict similar equilibrium bond lengths and conformational barriers.
As in the case of H$_2$, composite RHF/cc-pVTZ + CT-F12/q-UCCSD/cc-pVDZ energies leads to
equilibrium geometries and energy differences in agreement with CCSD/CBS.

\begin{table}[h!]
\centering
\resizebox{\columnwidth}{!}{
\begin{tabular}{cccccccccc}
\hline\hline
method/basis & type & $R^{tri}_{eq} [\text{\AA}]$ & $R^{lin}_{eq} [\text{\AA}]$ & $\Delta E$ [mHa] \\
\hline
RHF/6-31G                 & regular  & 0.843(1) & 0.798(1) & 53.8(2) \\
q-UCCSD/6-31G             & regular  & 0.855(1) & 0.809(1) & 49.9(2) \\
q-UCCSD/6-31G             & CT-F12   & 0.849(5) & 0.806(1) & 48.9(2) \\
\hline
RHF/cc-pVDZ               & regular  & 0.889(6) & 0.819(1) & 71.7(2) \\
q-UCCSD/cc-pVDZ           & regular  & 0.900(1) & 0.837(1) & 63.3(2) \\
q-UCCSD/cc-pVDZ           & CT-F12   & 0.895(1) & 0.832(1) & 64.6(2) \\
\hline
CCSD/CBS                & regular & 0.874(1)  & 0.814(1)  & 65.2(2) \\
cc-pVTZ/comp$^{(a)}$    & comp    & 0.874(1)  & 0.809(1)  & 67.6(2) \\
cc-pVTZ/comp$^{(b)}$    & comp    & 0.875(2)  & 0.816(1)  & 65.4(2) \\
\hline\hline
\end{tabular}
}
\caption{Equilibrium bond lengths for equilateral triangle and linear H$_3^+$, 
and energy difference between equilateral triangle and linear conformers. 
The listed quantities were obtained by locally fitting the computed potential energy surfaces to a Morse potential.
"comp" refers to the composite RHF/cc-pVTZ + CT-F12/q-UCCSD/6-31G 
and RHF/cc-pVTZ + CT-F12/q-UCCSD/cc-pVDZ methods (a,b respectively).
}
\label{tab:3}
\end{table}

\subsection{First-row hydrides}

In Sections \ref{sec:H2} and \ref{sec:H3plus} we explored hydrogen compounds. 
Here, we considered three closed-shell first-row hydrides: LiH, BH and HF.
We use RHF, q-UCCSD,and CT-F12/q-UCCSD with a 6-31G basis. 

\begin{figure}[h!]
\centering
\includegraphics[width=0.75\columnwidth]{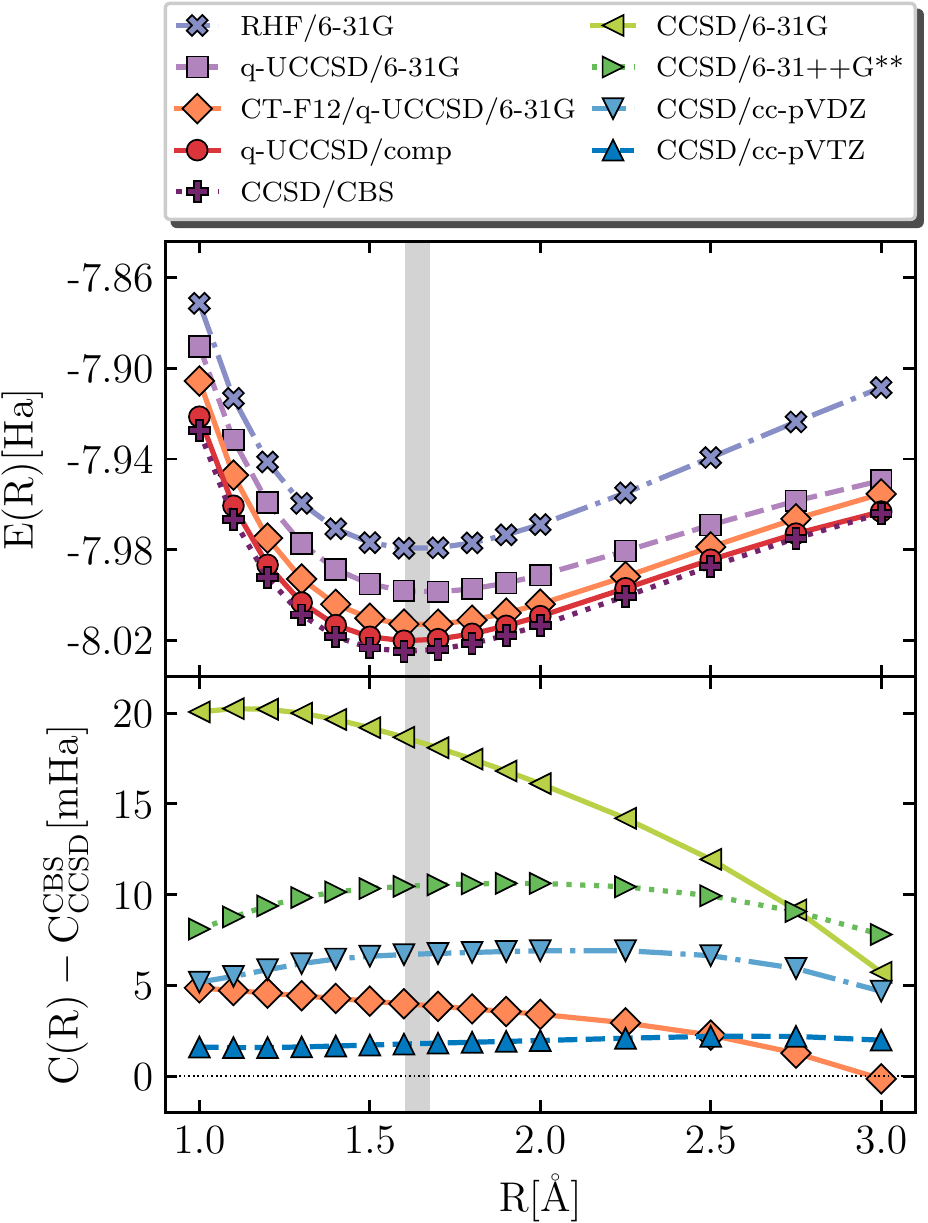}
\caption{
Top: Potential energy curves for LiH using RHF/6-31G, q-UCCSD/6-31G, CT-F12/q-UCCSD/6-31G, q-UCCSD/comp and CCSD/CBS.
Bottom: Comparison between classical CCSD/CBS correlation energies and classical CCSD/(6-31G, 6-31++G$^{**}$, cc-pVDZ, cc-pVTZ), CT-F12/q-UCCSD/6-31G correlation energies.
Lines are a guide for the eye, and gray bands represents the range of computed equilibrium bond lengths.
}
\label{fig:LiH}
\end{figure}

\begin{figure}[h!]
\centering
\includegraphics[width=0.75\columnwidth]{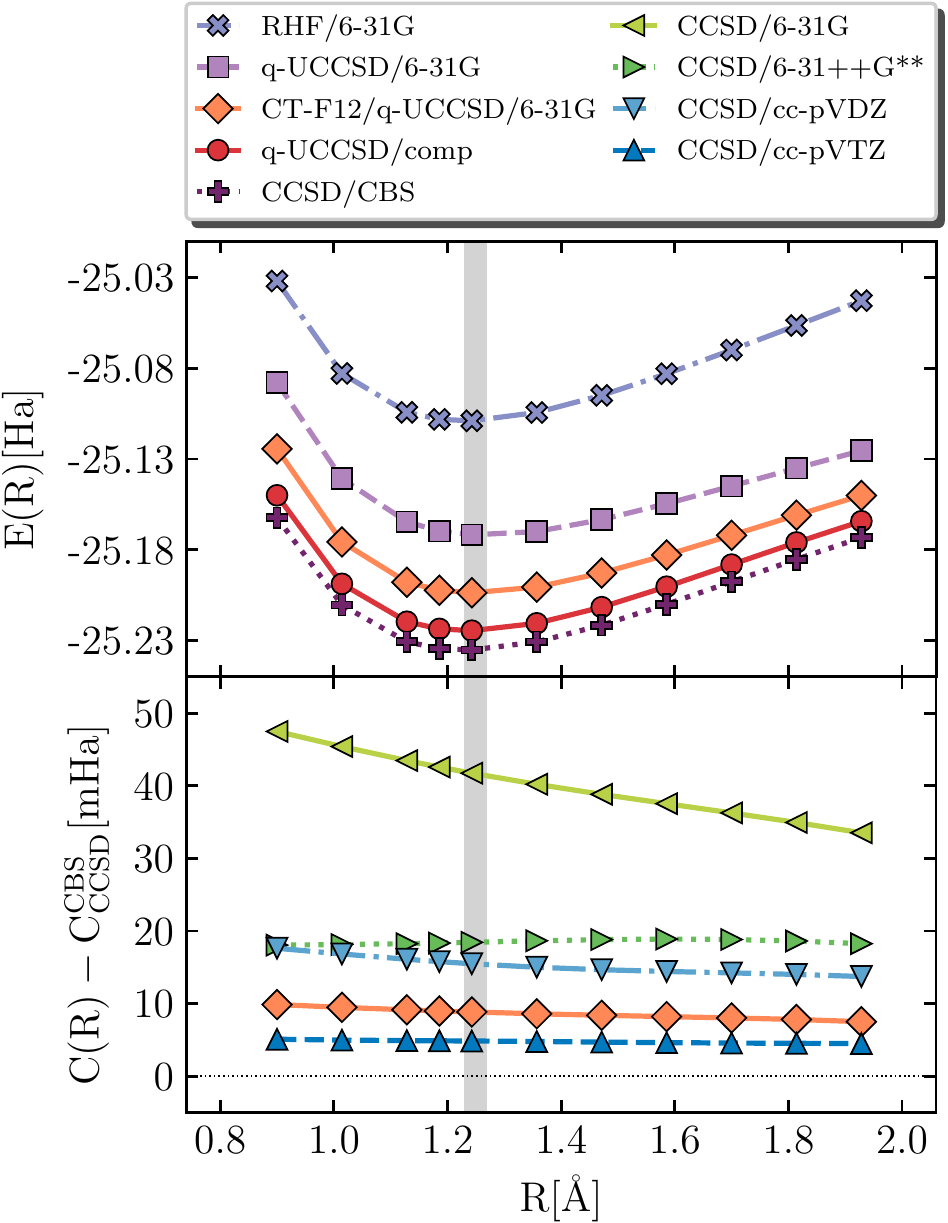}
\caption{
Top: Potential energy curves for BH using RHF/6-31G, q-UCCSD/6-31G, CT-F12/q-UCCSD/6-31G, q-UCCSD/comp and CCSD/CBS.
Bottom: Comparison between classical CCSD/CBS correlation energies and classical CCSD/(6-31G, 6-31++G$^{**}$, cc-pVDZ, cc-pVTZ), CT-F12/q-UCCSD/6-31G correlation energies.
Lines are a guide for the eye, and gray bands represents the range of computed equilibrium bond lengths.
}
\label{fig:BH}
\end{figure}

\begin{figure}[h!]
\centering
\includegraphics[width=0.75\columnwidth]{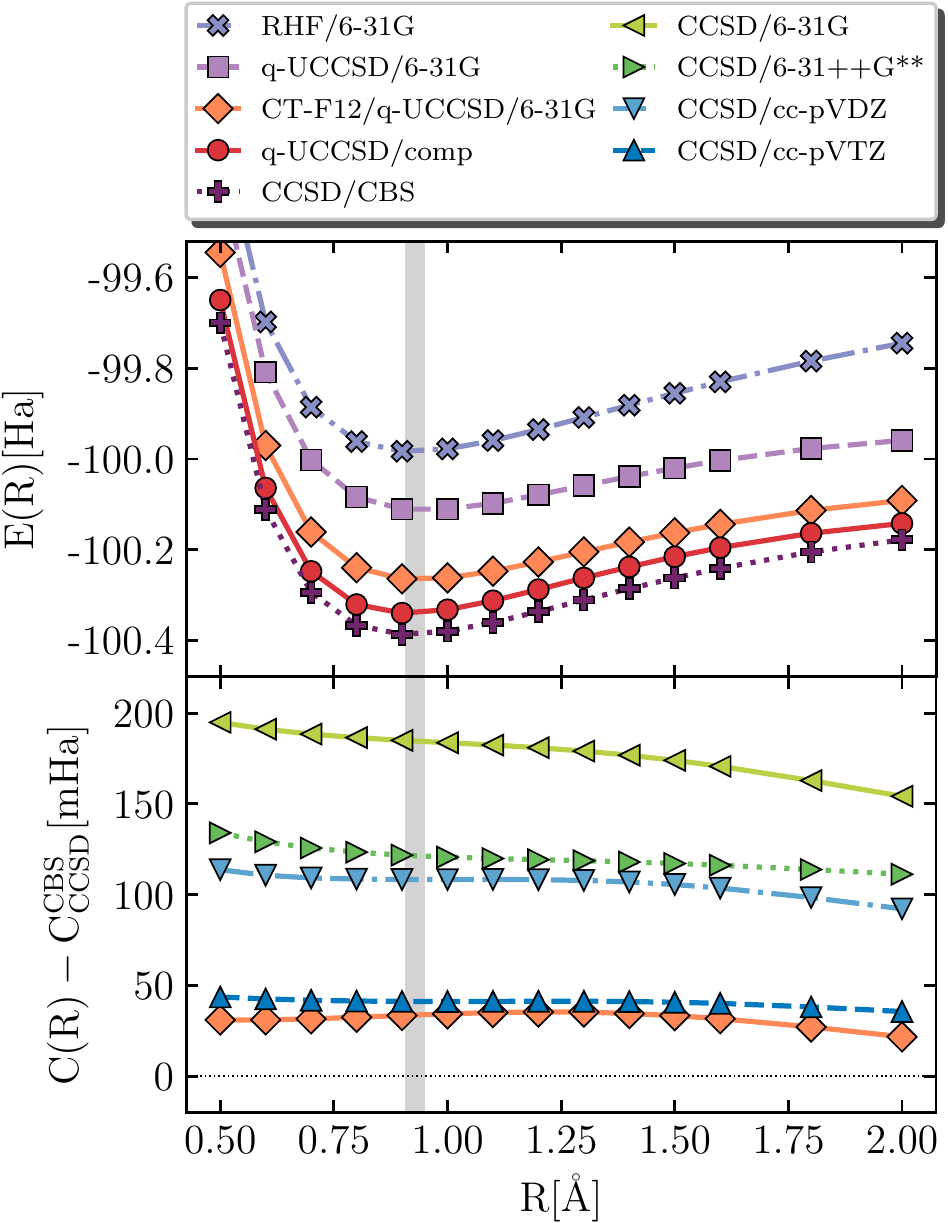}
\caption{
Top: Potential energy curves for HF using RHF/6-31G, q-UCCSD/6-31G, CT-F12/q-UCCSD/6-31G, q-UCCSD/comp and CCSD/CBS.
Bottom: Comparison between classical CCSD/CBS correlation energies and classical CCSD/(6-31G, 6-31++G$^{**}$, cc-pVDZ, cc-pVTZ), CT-F12/q-UCCSD/6-31G correlation energies.
Lines are a guide for the eye, and gray bands represents the range of computed equilibrium bond lengths.
}
\label{fig:HF}
\end{figure}

Results, including those with the composite method, are reported for LiH, BH and HF in Figures \ref{fig:LiH}, \ref{fig:BH} 
and \ref{fig:HF}, respectively. The trends observed for these molecules are again similar to those seen for H$_2$.
CT-F12/q-UCCSD/6-31G correlation energies have quality superior to CCSD/6-31++G$^{**}$ 
and CCSD/cc-pVDZ correlation energies, as shown in the bottom panels.

In Tables \ref{tab:LiH}, \ref{tab:BH} and \ref{tab:HF}, we list the results for 
equilibrium bond lengths and vibrational frequencies of LiH, BH and HF, respectively.
For all the hydrides considered here, CT-F12/q-UCCSD/6-31G geometries and frequencies
are closer to experimental and CCSD/CBS values than q-UCCSD/6-31G. For LiH and BH, vibrational frequencies further improve when the surface is described by the composite RHF/cc-pVTZ + CT-F12/q-UCCSD/6-31G energies. A similar effect is seen, in all species, for the equilibrium geometry.


\begin{table}[h!]
\centering
\begin{tabular}{cccccccccc}
\hline\hline
method/basis & type & $R_{eq} \, [\text{\AA}]$ & $\omega \, [cm^{-1}]$ \\
\hline
RHF/6-31G       & regular & 1.6369(1) & 1414(8) \\
q-UCCSD/6-31G   & regular & 1.6691(1) & 1287(8) \\
q-UCCSD/6-31G   & CT-F12  & 1.6477(1) & 1353(7) \\
q-UCCSD/comp    & comp    & 1.615(1)  & 1385(5) \\
CCSD/CBS        & regular & 1.607(2)  & 1390(5) \\
\hline\hline
\end{tabular}
\caption{Equilibrium bond length and vibrational frequencies for LiH, extracted from a 
Morse fit of potential energy curves. 
Experimental values are $R_{eq} = 1.595 \, \text{\AA}$ and $\omega = 1405 \, \mathrm{cm}^{-1}$, 
respectively \cite{johnson1999nist}.
The label ``comp" refers to the composite RHF/cc-pVTZ + CT-F12/q-UCCSD/6-31G method
}
\label{tab:LiH}
\end{table}

\begin{table}[h!]
\centering
\begin{tabular}{cccccccccc}
\hline\hline
method/basis & type & $R_{eq} \, [\text{\AA}]$ & $\omega \, [cm^{-1}]$ \\
\hline
RHF/6-31G             & regular  & 1.2328(7) & 2433(11) \\
q-UCCSD/6-31G         & regular  & 1.2671(5) &  2186(5) \\
q-UCCSD/6-31G         & CT-F12   & 1.2487(6) &  2287(7) \\
q-UCCSD/comp          & comp     & 1.232(1)  &  2364(7) \\
CCSD/CBS              & regular  & 1.234(1)  &  2369(5) \\
\hline\hline
\end{tabular}
\caption{
Equilibrium bond length and vibrational frequencies for BH, extracted from a 
Morse fit of potential energy curves.
Experimental values are $R_{eq} = 1.232 \, \text{\AA}$ and $\omega = 2367 \, \mathrm{cm}^{-1}$, 
respectively \cite{johnson1999nist}.
The label ``comp" refers to the composite RHF/cc-pVTZ + CT-F12/q-UCCSD/6-31G method
}
\label{tab:BH}
\end{table}

\begin{table}[h!]
\centering
\begin{tabular}{cccccccccc}
\hline\hline
method/basis & type & $R_{eq} \, [\text{\AA}]$ & $\omega \, [cm^{-1}]$ \\
\hline
RHF/6-31G     & regular & 0.920(2) & 4234(39) \\
q-UCCSD/6-31G & regular & 0.945(2) & 3836(33) \\
q-UCCSD/6-31G & CT-F12  & 0.935(2) & 3972(33) \\
cc-pVTZ/comp  & comp    & 0.910(1) & 4320(26) \\
CCSD/CBS      & regular & 0.913(1) & 4236(29) \\
\hline\hline
\end{tabular}
\caption{
Equilibrium bond length and vibrational frequencies for HF, extracted from a 
Morse fit of potential energy curves. 
Experimental values are $R_{eq} = 0.917 \, \text{\AA}$ and $\omega = 4138 \, \mathrm{cm}^{-1}$, 
respectively \cite{johnson1999nist}.
The label ``comp" refers to the composite RHF/cc-pVTZ + CT-F12/q-UCCSD/6-31G method
}
\label{tab:HF}
\end{table}

\subsection{Estimate of quantum resources}
\label{sec:D}
In the previous Sections, we explored energies, equilibrium geometries and vibrational properties 
of a collection of small molecules, assessing the accuracy of CT-F12/q-UCCSD.
In this Section, we estimate and compare the quantum resources needed to perform regular and explicitly correlated calculations for the chemical species considered in this work. 

\begin{table*}
\centering
\begin{tabular}{llcccrrrrr}
\hline\hline
system & basis & type & orbitals & qubits & Paulis$^{(a)}$ & parameters & operations & CNOTs & depth \\
\hline
H$_2$ & 6-31G                 & regular &  4 &  6 &   159 & 15 &   741 &   476 &   604 \\
H$_2$ & cc-pVDZ               & regular & 10 & 18 & 2,951 & 99 & 2,393 & 1,864 & 2,106 \\
H$_2$ & 6-31G                 & CT-F12  &  4 &  6 &   235 & 15 &   741 &   476 &   604 \\
H$_2$ & cc-pVDZ               & CT-F12  & 10 & 18 & 4,191 & 99 & 2,393 & 1,864 & 2,106 \\
\hline
H$_3^+$, triangular & 6-31G   & regular & 6 & 10 & 1,403 & 35 & 2,667 & 1,916 & 2,268 \\
%
H$_3^+$, triangular & cc-pVDZ & regular & 15 & 28 & 34,486 & 224 & 39,252 & 33,344 & 36,090 \\
%
H$_3^+$, triangular & 6-31G   & CT-F12 &  6 & 10 &  1,083 & 35 & 2,667 & 1,916 & 2,268 \\ 
%
H$_3^+$, triangular & cc-pVDZ & CT-F12 & 15 & 28 & 22,522 & 224 & 39,252 & 33,344 & 36,090 \\ 
\hline
LiH & 6-31G         & regular & 10 & 18 & 5,851 &  99 &  12,087 &  9,644 & 10,780 \\
LiH & 6-31G         & CT-F12  & 10 & 18 & 8,527 &  99 &  12,087 &  9,644 & 10,780 \\
\hline
BH & 6-31G          & regular & 10 & 18 & 5,851 & 344 & 44,087 & 35,180 & 37,241 \\ 
BH & 6-31G          & CT-F12  & 10 & 18 & 9,271 & 344 & 44,087 & 35,180 & 37,241 \\
\hline
HF & 6-31G          & regular & 10 & 18 & 5,851 & 804 & 104,027 & 82,628 & 86,120 \\
HF & 6-31G          & CT-F12  & 10 & 18 & 9,439 & 804 & 104,027 & 82,628 & 86,120 \\
\hline\hline
\end{tabular}
\caption{Columns 4-6: number of spatial orbitals, qubits and Pauli operators in the Hamiltonian 
for molecular species investigated in this work, at various levels of theory.
Columns 7-10: total number of parameters, quantum gates, CNOT gates and circuit depth in the VQE 
q-UCCSD and CT-F12/q-UCCSD circuits.
(a) matrix elements of the Hamiltonian smaller in absolute value than $10^{-8}$ Ha are truncated 
}
\label{tab:resources}
\end{table*}

The necessary quantum resources stem from the structure of the Hamiltonian operator and the VQE 
q-UCCSD circuit. Standard quantum encodings map the Fock space $\mathcal{F}_M$ of a molecular 
systems comprising $2M$ spin-orbitals onto the Hilbert space of $2M$ qubits,
\begin{equation}
\hat{\mathcal{E}} : \mathcal{F}_M \to \left( \mathbb{C}^2 \right)^{\otimes 2M}
\quad,\quad
\hat{\mathcal{E}} |x\rangle = | Ax \rangle
\quad,
\end{equation}
where $x \in \{0,1\}^{2M}$ is a binary string encoding a determinant, often with the convention
that the block of spin-up orbitals precedes the block of spin-down orbitals.
$A$ is an invertible $2M \times 2M$ binary matrix.
The standard Jordan-Wigner transformation is obtained by choosing $A$ as the identity matrix. 
The parity encoding instead uses
\begin{equation}
\begin{split}
A_0 &= 1 
\quad,\quad
A_1 = 
\left(
\begin{array}{cc}
1 & 0 \\
1 & 1 \\
\end{array}
\right)
\quad, \\
A_2 &= 
\left(
\begin{array}{cccc}
1 & 0 & 0 & 0 \\
1 & 1 & 0 & 0 \\
1 & 1 & 1 & 0 \\
1 & 1 & 1 & 1 \\
\end{array}
\right)
\quad \dots 
\end{split}
\end{equation}
As a result, for the parity encoding one has
\begin{equation}
\hat{\mathcal{E}} (-1)^{ \hat{N}_\uparrow} \hat{\mathcal{E}}^\dagger = \hat{Z}_M
\quad,\quad
\hat{\mathcal{E}} (-1)^{ \hat{N}_\uparrow + \hat{N}_\downarrow} \hat{\mathcal{E}}^\dagger = \hat{Z}_{2M}
\quad,
\end{equation}
where $Z_i$ denotes the Pauli Z operator acting on qubit $i$.

Conservation of spin-up and spin-down particle numbers modulo 2 can be enforced by freezing 
qubits $M$ and $2M$ in eigenvectors of $Z_M$ and $Z_{2M}$ with suitable eigenvalues, thereby 
reducing the number of qubits by 2.

A similar reduction of qubits can be achieved in presence of point-group $\mathbb{Z}_2$ 
symmetries. Denoting $\{ \hat{\tau}_i \}_{i=1}^k$ the generators of the Hamiltonian symmetry group,
it can be proved \cite{bravyi2017tapering,setia2019reducing} that there exists a Clifford 
transformation $\hat{U}$, computable at polynomial cost on a conventional computer, such that
\begin{equation}
\hat{U} \hat{\mathcal{E}} \hat{\tau}_i \hat{\mathcal{E}}^\dagger \hat{U}^\dagger 
= \hat{X}_{i}
\quad.
\end{equation}
The simulation can thus be restricted to an irreducible representation of the $\mathbb{Z}_2$ 
symmetry under consideration by freezing qubit $i$ into an eigenvector of $X_i$.

In combination with the parity encoding \cite{seeley2012bravyi,tranter2018comparison}, 
conservation of spin-up and spin-down particle numbers reduces the number of qubits by 2, 
and tapering off techniques can be used to bring the number of qubits to $N_q = 2M-2-k$.

\begin{table*}
\centering
\begin{tabular}{llccrrrr}
\hline\hline
system  & basis   & orbitals & qubits & parameters & operations & CNOTs & depth \\
\hline
H$_2$   & 6-31G   &  4 &  8 &      15 &      1,478 &        768 &        979 \\
H$_2$   & cc-pVDZ & 10 & 20 &      99 &     20,630 &     14,616 &     16,435 \\
H$_2$   & cc-pVTZ & 28 & 56 &     783 &    394,310 &    341,280 &    357,427 \\
\hline
H$^+_3$ & 6-31G   &  6 & 12 &      35 &      4,822 &      2,920 &      3,491 \\
H$^+_3$ & cc-pVDZ & 15 & 30 &     224 &     65,410 &     51,016 &     55,385 \\
H$^+_3$ & cc-pVTZ & 42 & 84 &   1,763 &  1,285,270 &  1,163,416 &  1,200,563 \\
\hline
LiH     & 6-31G   & 10 & 20 &      99 &     20,630 &     14,616 &     16,435 \\
LiH     & cc-pVDZ & 18 & 36 &     323 &    110,230 &     89,080 &     95,507 \\
LiH     & cc-pVTZ & 43 & 86 &   1,848 &  1,376,930 &  1,249,080 &  1,288,057 \\
\hline
BH      & 6-31G   & 10 & 20 &     344 &     72,964 &     50,176 &     54,529 \\
BH      & cc-pVDZ & 18 & 36 &   1,328 &    434,692 &    343,040 &    354,817 \\
BH      & cc-pVTZ & 43 & 86 &   8,528 &  5,771,492 &  5,167,640 &  5,132,217 \\
\hline
HF      & 6-31G   & 10 & 20 &    804 &     171,656 &    116,736 &    125,185 \\
HF      & cc-pVDZ & 18 & 36 &  4,340 &   1,396,872 &  1,091,328 &  1,111,041 \\
HF      & cc-pVTZ & 43 & 86 & 33,540 &  21,831,272 & 19,435,728 & 18,975,841 \\
\hline\hline
\end{tabular}
\caption{Number of orbitals, qubits and number of parameters, operations, CNOTs and depth of the 
q-UCCSD and CT-F12/q-UCCSD circuits for various systems, for the species studied in this work.
The Jordan-Wigner mapping and frozen core approximation (for Li, B, F) were used, without 
truncations of small terms or circuit transpilation
}
\label{tab:compare}
\end{table*}

Under the chosen encoding, and in presence of tapering techniques, the Hamiltonian takes the form
\begin{equation}
\label{eq:ham_qubit}
\hat{H} = \sum_{i=1}^{N_p} c_i \hat{P}_i \quad,
\end{equation}
where $\hat{P}_i$ is a tensor product of $N_q$ Pauli operators,
\begin{equation}
\hat{P}_i = \hat{\sigma}_{i_1} \otimes \dots \otimes \hat{\sigma}_{i_{N_q}} \in \{ \hat{I}, \hat{X}, \hat{Y}, \hat{Z} \}^{N_q} \;,
\end{equation}
where $\hat{X}, \hat{Y}, \hat{Z}$ denote the spin-$\frac{1}{2}$ Pauli operators.
Naturally, the number $N_p$ of terms in Eq.~\eqref{eq:ham_qubit} is an important quantum resource, 
because it affects the number of measurements needed to estimate the expectation value of $\hat{H}$.

The q-UCCSD and CT-F12/q-UCCSD circuits can be implemented by a Trotter decomposition,
\begin{equation}
\label{eq:vqe_uccsd}
\hat{U}(\theta) \simeq \left[
\prod_{ia} 
e^{ \frac{ \theta^a_i }{N_s} 
\left( \hat{c}^\dagger_a \hat{c}_i 
- 
\hat{c}^\dagger_i \hat{c}_a \right)}
\prod_{ijab} 
e^{ \frac{ \theta^{ab}_{ij} }{N_s} 
\left( \hat{c}^\dagger_a \hat{c}^\dagger_b \hat{c}_j \hat{c}_i 
- 
\hat{c}^\dagger_i \hat{c}^\dagger_j \hat{c}_b \hat{c}_a \right)}
\right]^{N_s}
\end{equation}
where $N_s$ is the number of slices in a Trotter implementation of the q-UCCSD or CT-F12/q-UCCSD 
operator. In this work, we used $N_s = 1$ time slices in all calculations, and a first-order Trotter scheme with two-body and one-body excitations applied consecutively. Unlike $e^{ \hat{T} - \hat{T}^\dagger }$, each of the exponentials in the right-hand 
side of Eq.~\eqref{eq:vqe_uccsd} can be mapped onto a circuit comprising a number of single-qubit 
and CNOT gates that scale at most linearly with the number of qubits $N_q$. 

It is worth pointing out that the description given in this section refers to the implementation of q-UCCSD in the Qiskit package. In recent times, a number of methodological developments have given rise to implementations with lower gate complexity, for example through low-rank decompositions and recompilation techniques \cite{motta2018low,motta2019efficient,matsuzawa2020jastrow,cowtan2020generic,xia2020coupled}.
Similarly, the impact of Trotterization 
\cite{evangelista2019exact,gard2020efficient,grimsley2019trotterized}
has been understood more profoundly and established more firmly.
In this work, we made the operational decision
to integrate CT-F12 into an existing and publicly available computational package for q-UCCSD calculation. Exploration of more efficient strategies and extension to other algorithms and variational forms are important topics, that should be addressed in future research.

To characterize the computational cost of a q-UCCSD or CT-F12/q-UCCSD simulation, it is 
important to know the number of parameters $\theta$ to be optimized, the number of quantum 
operations (one- and two-qubit gates) and especially CNOT gates comprising the circuit 
$\hat{U}(\theta)$, and the circuit depth, corresponding to the number of groups of quantum 
gates that cannot be executed in parallel. Of course, circuits comprising more gates, especially
CNOT gates, and featuring higher depth, are more expensive.

We list all these parameters in Table \ref{tab:resources}. To reduce the number of qubits, 
we used $\mathbb{Z}_2$ symmetries that conserve the number of spin-up and spin-down particles.
An important and encouraging observation is that the cost of an explicitly 
correlated calculation with underlying basis $B$, for example, CT-F12/q-UCCSD/6-31G, is 
essentially identical to that of a regular simulation with underlying basis $B$, q-UCCSD/6-31G. 
The only difference is represented by the higher number of Pauli operators in the Hamiltonian, 
which in turn is due to the loss of 8-fold symmetry in favor of 4-fold symmetry. 
In fact, the number of Pauli operators in the Hamiltonian is dominated by the two-body contribution, due to the summation over the $N^4$ elements of the electron repulsion integral $(pr|qs)$. This summation reduces to $N^2 (N+1)^2 / 4$ terms in presence of 4-fold symmetry
$(pr|qs) = (rp|sq) = (qs|pr)$, and to $ N^2 (N+1)^2 / 8$ terms in presence of 8-fold symmetry, $(pr|qs) = (rp|sq) = (qs|pr)$ and $(pr|qs) = (rp|qs)$ \cite{sun2018pyscf,sun2020recent}.
This is why the number of Pauli operators in the CT-F12 Hamiltonian is roughly twice that of the regular Hamiltonian. Other differences seen in Table \ref{tab:resources} are due to the one-body Hamiltonian, truncation thresholds and molecular symmetries.

Despite the higher number of Pauli operator, and much more importantly, a CT-F12/q-UCCSD/$B$ calculation (here $B$ denotes the underlying basis) 
yields results of accuracy comparable with those from a q-UCCSD/$B^\prime$ with $B^\prime$ 
larger than $B$, which can result in a quantum simulation \ita{ several orders of magnitude 
more expensive.}
Table \ref{tab:compare} lists a number of properties to consider before performing q-UCCSD/(6-31G, 
cc-pVDZ, cc-pVTZ) calculations for the systems considered in this work. 
The numbers quoted in Table \ref{tab:compare} provide an estimate of the quantum resources needed 
to carry out such simulations, rather than their precise requirements. 
This is meant to help appreciate how CT-F12 economizes q-UCCSD simulations.
For example, the qubits required by a cc-pVTZ simulations is roughly 4 times that required by
a 6-31G simulation. Similarly, the number of CNOT gates in a q-UCCSD/cc-pVTZ circuit is roughly
2 orders of magnitude higher than the corresponding one with a 6-31G basis set.
We emphasise that the reduction in CNOT gates observed here arises primarily from the use of transcorrelation: a 6-31G basis and a transcorrelated Hamiltonian are equivalent in accuracy to a cc-pVTZ basis and a standard Hamiltonian. Since the former basis is more compact, i.e. it has less orbitals, any calculation performed with it requires less qubits and gates, in the amount specified above.

It is reasonable to assume that CT-F12/q-UCCSD/6-31G provide correlation energies comparable to 
q-UCCSD/cc-pVTZ correlation energies, since composite methods yield potential energy curves of 
quality near to CCSD/cc-pVTZ. For example, see Figures \ref{fig:1}, \ref{fig:BH} and \ref{fig:HF}, 
where the composite RHF/cc-pVTZ + CT-F12/q-UCCSD/6-31G (RHF/cc-pVTZ + CT-F12/q-UCCSD/cc-pvDZ for 
$H_2$) curves lie almost on top of CCSD/cc-pVTZ curves.

\section{Conclusions}
\label{sec:conclusions}

To increase the accuracy of quantum simulations of chemical systems,
we explored the use of \ita{ab initio} Hamiltonians similarity-transformed to incorporate
dynamical electron correlation effects. 

Our work takes a step towards removing an important limitation of quantum simulations of chemical systems, namely the low quality of energies and properties resulting from the use of minimal basis sets. For the molecular species we studied, the number of qubits needed to simulate a 6-31G basis yielded energies and properties of cc-pVTZ quality. 

Other favorable traits of the similarity-transformed Hamiltonian considered here (CT-F12 Hamiltonian) include its hermiticity, absence of 
two-electron singularities, and inclusion of only one- and two-body operators.

The improvement in the accuracy of energies and properties requires only  a very modest increase in the necessary quantum resources, when compared to regular (non CT-F12) calculations with the same basis set. In particular, the increase is limited to the number of Pauli operators in the qubit representation of the Hamiltonian. 

We elected to focus specifically on the CT-F12 method and so we used the q-UCCSD algorithm due to its widespread use in published literature and computational packages. 
Nevertheless, the results obtained here will straightforwardly translate to many other quantum algorithms for quantum chemistry. Examples of such algorithms include quantum subspace expansion \cite{mcclean_pra_2017}, quantum equation of motion \cite{ollitrault2019quantum} and quantum phase estimation \cite{kitaev1995quantum}.
Although we demonstrated a dramatic reduction in the quantum resources required by CT-F12 
q-UCCSD simulations, this algorithm still far exceeds the budget of contemporary quantum hardware in terms of both entangling gates and circuit depth due to the use of the q-UCCSD Ansatz. Research into the combination of CT-F12 techniques and hardware-efficient Ans\"{a}tze, that can be demonstrated on contemporary quantum hardware, is underway.

\section*{Conflicts of interest}

There are no conflicts to declare.

\section*{Acknowledgements}

TG, JL, MM and JER acknowledge the IBM Research Cognitive Computing Cluster service for 
providing resources that have contributed to the research results reported within this 
paper. The work of AK, CM, and EFV was supported by the U.S. National Science Foundation 
(awards 1550456 and 1800348)

\providecommand*{\mcitethebibliography}{\thebibliography}
\csname @ifundefined\endcsname{endmcitethebibliography}
{\let\endmcitethebibliography\endthebibliography}{}

\end{document}